\documentclass[twocolumn,prl,superscriptaddress,showpacs,epsf]{revtex4}

\usepackage{graphicx}

\begin{document}
\title{Scalable quantum computing with Josephson charge qubits}
%
%
\date{\today}
\author{J. Q. You}
\affiliation{Frontier Research System, The Institute of Physical
and Chemical Research (RIKEN), Wako-shi 351-0198, Japan}
\author{J. S. Tsai}
\affiliation{Frontier Research System, The Institute of Physical
and Chemical Research (RIKEN), Wako-shi 351-0198, Japan}
\affiliation{NEC Fundamental Research Laboratories, Tsukuba,
Ibaraki 305-8051, Japan} \altaffiliation[Permanent address.]{}
\author{Franco Nori}
\altaffiliation[Corresponding author~(Email:~nori@umich.edu).]{}
\affiliation{Frontier Research System, The Institute of Physical
and Chemical Research (RIKEN), Wako-shi 351-0198, Japan}
\affiliation{Center for Theoretical Physics, Physics Department,
Center for the Study of Complex Systems, The University of
Michigan, Ann Arbor, MI 48109-1120, USA} \altaffiliation[Permanent
address.]{}

\begin{abstract}
A goal of quantum information technology is to control the quantum state
of a system, including its preparation, manipulation, and measurement.
However, scalability to many qubits and controlled connectivity between any selected qubits
are two of the major stumbling blocks to achieve quantum computing (QC). Here we propose
an experimental method, using Josephson charge qubits,
to efficiently solve these two central problems.
The proposed QC architecture is {\it scalable}
since any two charge qubits can be effectively coupled by an experimentally accessible
inductance. More importantly, we formulate an {\it efficient} and {\it realizable} QC scheme
that requires {\it only one} (instead of two or more) two-bit operation to implement
conditional gates.
\end{abstract}
\maketitle

The macroscopic quantum effects in low-capacitance Josephson-junction circuits
have recently been used to realize qubits for quantum information processing,
and these qubits are expected to be scalable to large-scale circuits using
modern microfabrication techniques.
Josephson-qubit devices~\cite{MSS} are based on the charge and phase degrees
of freedom. The charge qubit is achieved in a Cooper-pair box~\cite{NPT},
where two dominant charge states are coupled through coherent Cooper-pair
tunneling~\cite{ASZ}, while the phase qubit is based on two different
flux states in a small superconducting-quantum-interference-device
(SQUID) loop~\cite{MOOIJ,ORLANDO}.
Experimentally, the energy-level
splitting and the related properties of state superpositions were
observed via Cooper-pair tunneling in the Josephson charge
device~\cite{NCT,BOUCH} and by spectroscopic measurements for
the Josephson phase device~\cite{VAL,FRIED}. Moreover,
coherent oscillations were demonstrated in a Josephson
charge device prepared in a superposition of two charge
states~\cite{NPT}. These striking experimental observations reveal
that the Josephson charge and phase devices are suitable for
solid-state qubits in quantum information processing.
The next immediate challenge would include implementing a
{\it two-bit coupling} and then {\it scaling up} the architecture
to many qubits. Here, we focus on the Josephson charge qubit realized in a
Cooper-pair box and propose a new quantum-computing (QC) scheme
based on scalable charge-qubit structures.

A straightforward way of coupling Josephson charge qubits is to
use the Coulomb interactions between charges on different islands
of the charge qubits ({\it e.g.}, to connect two Cooper-pair boxes
via a capacitor). A two-bit operation~\cite{PFP}, similar to the
controlled-NOT gate, was derived using this interbit coupling; but
it is hard to switch the coupling on and off~\cite{MSS} in this
scheme as well as to make the system scalable because only
neighboring qubits can be coupled. A scalable way of coupling
Josephson charge qubits was proposed~\cite{MSS,ASZ} in terms of
the oscillator modes in a $LC$ circuit formed by an inductance and
the qubit capacitors. In this design, the interbit coupling is
switchable and any two charge qubits can be coupled.
%
%
However, there is no efficient (i.e., using one two-bit operation)
QC scheme for this design~\cite{MSS,ASZ} to achieve conditional
gates such as the controlled-phase-shift and controlled-NOT gates.
Moreover, the calculated interbit coupling terms~\cite{MSS,ASZ}
only apply to the case when two conditions are met: (i) the
eigenfrequency $\omega_{LC}$ of the $LC$ circuit is much faster than the 
quantum manipulation frequencies ({\it which limits the allowed number $N$
of the qubits in the circuit} because $\omega_{LC}$ scales with $1/\sqrt{N}$)   
and (ii) the phase conjugate to the total
charge on the qubit capacitors fluctuates weakly.
These two limitations do not apply to our approach.
In our proposal, a common inductance (but no $LC$ circuit)
is used to couple all Josephson charge qubits.
In our scheme, both dc and ac supercurrents can flow through the
inductance, while in~\cite{MSS,ASZ} only ac supercurrents can flow
through the inductance and it is the $LC$-oscillator mode that
couples the charge qubits.
These yield different interbit couplings ({\it e.g.}, $\sigma_y
\sigma_y$ type~\cite{MSS,ASZ} as opposed to  $\sigma_x \sigma_x$
in our scheme).
To have a {\it controllable interbit coupling}, we employ two dc
SQUIDs to connect each Cooper-pair box. Our proposed QC
architecture is scalable in the sense that {\it any \/} two charge
qubits ({\it not \/} necessarily neighbors) can be effectively
coupled by an experimentally accessible inductance. More
importantly, we formulate an efficient QC scheme that requires
{\it only one} (instead of two or more) two-bit operation to
implement conditional gates. To our knowledge, this is the first
efficient scalable QC scheme for this type of architecture.

\begin{figure}
\includegraphics[width=3.3in,height=1.7in,
bbllx=45,bblly=410,bburx=540,bbury=670]{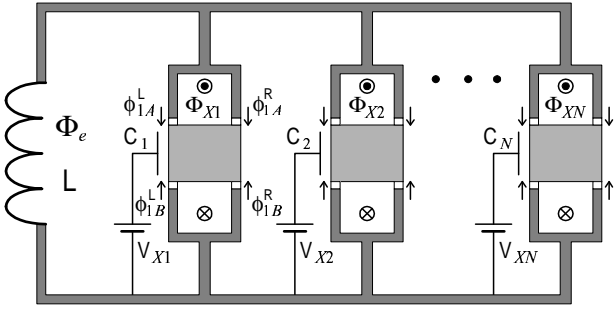}
\caption{Schematic diagram of the proposed scalable and switchable
quantum computer.
%
%
%
All Josephson charge-qubit structures are coupled by a common
superconducting inductance.
Here, each Cooper-pair box is operated both in the charging regime
$E_{ck}\gg E^0_{Jk}$ and at low temperatures $k_BT\ll E_{ck}$.
Moreover, the superconducting gap is larger than $E_{ck}$, so that
quasiparticle tunneling is prohibited in the system. }
\label{fig1}
\end{figure}

The proposed quantum computer consists of $N$ Cooper-pair boxes coupled by a common
superconducting inductance $L$ (see Fig.~1). For the $k$th Cooper-pair box, a superconducting
island with charge $Q_k=2en_k$ is weakly coupled by two symmetric dc SQUIDs and biased by an
applied voltage $V_{Xk}$ through a gate capacitance $C_k$. The two symmetric dc SQUIDs are
assumed to be identical and all Josephson junctions in them have Josephson coupling energy
$E^0_{Jk}$ and capacitance $C_{Jk}$.
Each SQUID pierced by a magnetic flux $\Phi_{Xk}$
provides an effective coupling energy given by
$-E_{Jk}(\Phi_{Xk})\cos\phi_{kA(B)}$; with
$E_{Jk}(\Phi_{Xk})=2E^0_{Jk}\cos(\pi\Phi_{Xk}/\Phi_0)$,
and $\Phi_0=h/2e$ is the flux quantum.
The effective phase drop $\phi_{kA(B)}$,
with subscript $A(B)$ labelling the SQUID above (below) the island,
equals the average value,
$[\phi^L_{kA(B)}+\phi^R_{kA(B)}]/2$, of the phase drops
across the two Josephson junctions in the dc SQUID;
where the superscript $L$ ($R$) denotes the left (right) Josephson junction.
Since the size of the loop is usually very small ($\sim 1$ $\mu$m),
above we have ignored the self-inductance effects of each SQUID loop.
The Hamiltonian of the system is
$ H=\sum_{k=1}^N H_k+{1\over 2}LI^2 \, $,
with $H_k$ given by
$ H_k=E_{ck}(n_k-n_{Xk})^2-E_{Jk}(\Phi_{Xk})(\cos\phi_{kA}+\cos\phi_{kB}). $
%
Here, $E_{ck}=2e^2/(C_k+4C_{Jk})$ is the charging energy of the superconducting island and
$I=\sum_{k=1}^NI_k$ is the total persistent current through the superconducting inductance,
as contributed by all coupled Cooper-pair boxes. 
The offset charge $2en_{Xk}=C_kV_{Xk}$ is induced
by the gate voltage $V_{Xk}$. The phase drops $\phi^L_{kA}$
and $\phi^L_{kB}$ are related to the total flux $\Phi=\Phi_e+LI$ through the inductance
$L$ by the constraint $\phi^L_{kB}-\phi^L_{kA}=2\pi\Phi/\Phi_0$, where $\Phi_e$ is
the externally applied magnetic flux threading the inductance $L$.
Without loss of generality and in order to implement QC more
conveniently, the magnetic fluxes through the two SQUID loops of
each Cooper-pair box are designed to have the {\it same} values
but {\it opposite} directions; this simplifies the form of the
Hamiltonian. (If this were not to be the case, the interbit
coupling can still be realized, but the Hamiltonian of the qubit
circuits takes a more complicated form.)
Because this pair of fluxes {\it cancel} each other in any loop enclosing them,
then $\phi^L_{kB}-\phi^L_{kA}=\phi^R_{kB}-\phi^R_{kA}$.
This gives rise to the constraint $\phi_{kB}-\phi_{kA}=2\pi\Phi/\Phi_0$
for the average phase drops across the Josephson junctions
in the SQUIDs.
The common superconducting inductance $L$ plays the role of coupling
Cooper-pair boxes. The coupling of selected Cooper-pair boxes can be implemented
by switching on the SQUIDs connected to the chosen Cooper-pair boxes,
and the persistent currents through the inductance $L$ are composed of
contributions from all the coupled Cooper-pair boxes.

\begin{figure}
\includegraphics[width=2.2in,height=1.3in,bbllx=12,bblly=480,bburx=230,bbury=630]{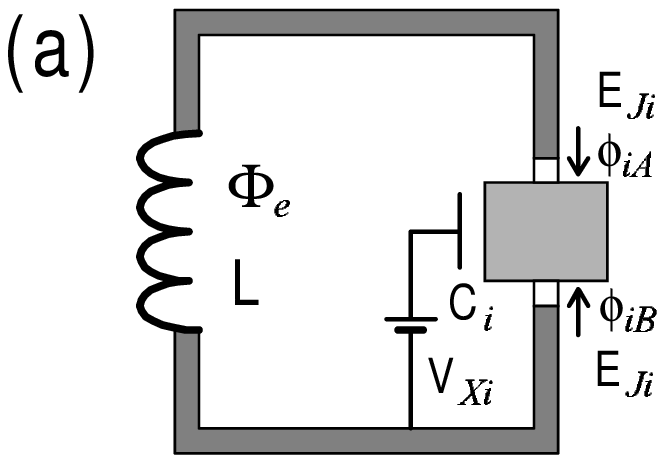}
\includegraphics[width=2.7in,height=1.3in,bbllx=240,bblly=485,bburx=530,bbury=635]{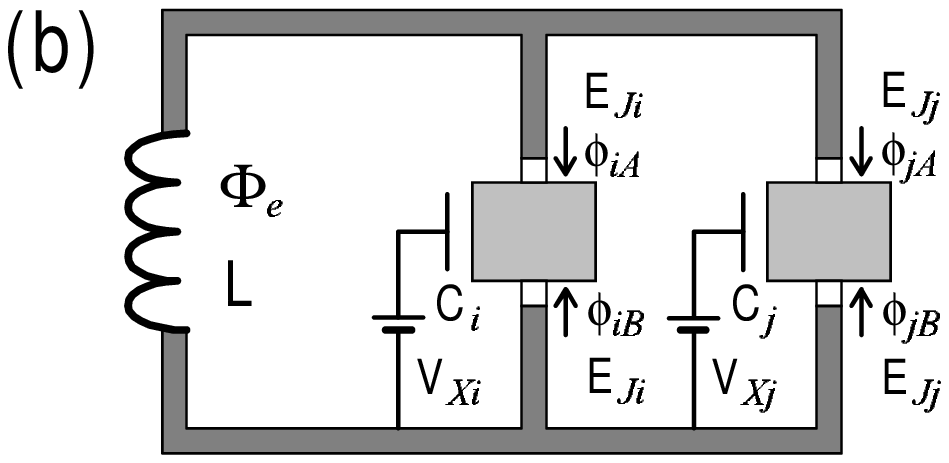}
\caption{(a) One-bit circuit with a Cooper-pair box connected to
the inductance. (b) Two-bit structure where two Cooper-pair boxes
are commonly connected to the inductance. Here, each SQUID
connecting the superconducting island is represented by an
effective Josephson junction.} \label{fig2}
\end{figure}

{\it One- and two-bit circuits.}~{\bf ---}~For any given Cooper-pair box, say $i$,
when $\Phi_{Xk}={1\over 2}\Phi_0$ and $V_{Xk}=(2n_k+1)e/C_k$ for all boxes except $k=i$,
the inductance $L$ only connects the $i$th Cooper-pair box to form a superconducting
loop [see Fig.~2(a)]. The Hamiltonian of the system can be reduced to~\cite{eqn1}
\begin{equation}
H=\varepsilon_i(V_{Xi})\,\sigma_z^{(i)}-\overline{E}_{Ji}(\Phi_{Xi},\Phi_e,L)
\;\sigma_x^{(i)},
\end{equation}
where $\varepsilon_i(V_{Xi})$ is controllable via the gate voltage
$V_{Xi}$, while the intrabit coupling
$\overline{E}_{Ji}(\Phi_{Xi},\Phi_e,L)$
can be controlled by both the applied external flux $\Phi_e$
through the common inductance, and the local flux $\Phi_{Xi}$
through the two SQUID loops of the i-th Cooper-pair box.
The intrabit coupling $\overline{E}_{Ji}$ in (1) is different from
that in~\cite{MSS,ASZ} because a very different contribution by
$L$ is considered.
To couple {\it any} two Cooper-pair boxes, say $i$ and $j$,
we choose $\Phi_{Xk}={1\over 2}\Phi_0$ and $V_{Xk}=(2n_k+1)e/C_k$ for all boxes
except $k=i$ and $j$. As shown in Fig.~2(b), the inductance $L$ is shared by the
Cooper-pair boxes $i$ and $j$ to form superconducting loops. The reduced Hamiltonian
of the system is given by~\cite{eqn2}
\begin{equation}
H=\sum_{k=i,j}[\varepsilon_k(V_{Xk})\,\sigma_z^{(k)}-\overline{E}_{Jk}\;\sigma_x^{(k)}]
+\Pi_{ij}\,\sigma^{(i)}_x\sigma^{(j)}_x.
\end{equation}
Here the interbit coupling $\Pi_{ij}$ is controlled by both the
external flux $\Phi_e$ through the inductance $L$, and the local
fluxes, $\Phi_{Xi}$ and $\Phi_{Xj}$, through the SQUID loops.
%


{\it Quantum computing.}~{\bf ---}~The
quantum system evolves according to $U(t)=\exp(-iHt/\hbar)$.
Initially, we choose $\Phi_{Xk}={1\over 2}\Phi_0$ and $V_{Xk}=(2n_k+1)e/C_k$
for all boxes in Fig.~1, so that the Hamiltonian of the system is $H=0$
and no time evolution occurs.
Afterwards, we {\it switch} certain fluxes $\Phi_{Xk}$ and/or gate voltages $V_{Xk}$
away from the above initial values for certain periods of times,
to implement logic gates required for QC.
For any two Cooper-pair boxes, say $i$ and $j$, when fluxes
$\Phi_{Xi}$ and $\Phi_{Xj}$ are switched away from the initial value $\Phi_0/2$ for a given
period of time $\tau$, the Hamiltonian of the system becomes
$H=-\overline{E}_{Ji}\sigma_x^{(i)}-\overline{E}_{Jj}\sigma_x^{(j)}
+\Pi_{ij}\sigma^{(i)}_x\sigma^{(j)}_x$. This anisotropic Hamiltonian is Ising-like~\cite{BURK},
with its anisotropic direction and the ``magnetic" field along the $x$ axis.
When the parameters are suitably chosen so that
$\overline{E}_{Ji}=\overline{E}_{Jj}=\Pi_{ij}=-\pi\hbar/4\tau$ for the switching time
$\tau$, we obtain a controlled-phase-shift gate:
$U'_{\rm CPS}=e^{i\pi/4}U_{2b}
=\exp\left\{i{\pi\over 4}[1-\sigma^{(i)}_x-\sigma^{(j)}_x+\sigma^{(i)}_x\sigma^{(j)}_x]\right\}$,
which does not alter the two-bit states $|+\rangle_i|+\rangle_j$, $|+\rangle_i|-\rangle_j$,
and $|-\rangle_i|+\rangle_j$, but transforms
$|-\rangle_i|-\rangle_j$ to $-|-\rangle_i|-\rangle_j$.
Here, the phase factor $e^{i\pi/4}$ corresponds to an overal energy shift of the Hamiltonian,
and $|\pm\rangle$ are defined by
$|\pm\rangle=(|\!\uparrow\rangle\pm |\!\downarrow\rangle)/\sqrt{2}$.


To obtain the controlled-phase-shift gate $U_{\rm CPS}$ for the basis states
$|\!\uparrow\rangle_i|\!\uparrow\rangle_j$, $|\!\uparrow\rangle_i|\!\downarrow\rangle_j$,
$|\!\downarrow\rangle_i|\!\uparrow\rangle_j$, and $|\!\downarrow\rangle_i|\!\downarrow\rangle_j$,
one needs to combine $U'_{\rm CPS}$ with suitable one-bit rotations.
For any Cooper-pair box, say $i$, one can shift flux $\Phi_{Xi}$ and/or gate
voltage $V_{Xi}$ for a given switching time $\tau$ to derive one-bit rotations.
A universal set of one-bit gates $U^{(i)}_z(\alpha)=e^{i\alpha\sigma^{(i)}_z}$, and
$U^{(i)}_x(\beta)=e^{i\beta\sigma^{(i)}_x}$, where $\alpha=-\varepsilon_i(V_{Xi})\tau/\hbar$
and $\beta=\overline{E}_{Ji}\tau/\hbar$, can be defined by choosing
$\overline{E}_{Ji}=0$ and $\varepsilon_i(V_{Xi})=0$ (which can be done with
suitable choices of $\Phi_{Xi}$ and $V_{Xi}$) in the one-bit Hamiltonian (1),
respectively. Any one-bit rotation can be derived in terms of these two types of one-bit gates.
For instance, the Hadamard gate is given by
${\cal H}_i=e^{-i\pi/2}U^{(i)}_z({\pi\over 4})U^{(i)}_x({\pi\over 4})U^{(i)}_z({\pi\over 4})$.
Using ${\cal H}_i$, we derive the controlled-phase-shift gate $U_{\rm CPS}$:
$ \; U_{\rm CPS}={\cal H}^{\dag}_j{\cal H}^{\dag}_iU'_{\rm
CPS}{\cal H}_i{\cal H}_j $.
The one-bit rotation
$V_j=e^{i\pi\sigma^{(j)}_y/4}$ is given by
$V_j=U^{(j)}_z(-{\pi\over 4})U^{(j)}_x({\pi\over 4})U^{(j)}_z({\pi\over 4})$.
Combining $V_j$ with $U_{\rm CPS}$, we obtain the controlled-NOT gate:
$\; U_{\rm CNOT}=V^{\dag}_jU_{\rm CPS}V_j \, $,
which transforms the basis states as
$|\!\uparrow\rangle_i|\!\uparrow\rangle_j\longrightarrow|\!\uparrow\rangle_i|\!\uparrow\rangle_j$,
$|\!\uparrow\rangle_i|\!\downarrow\rangle_j
\longrightarrow|\!\uparrow\rangle_i|\!\downarrow\rangle_j$,
$|\!\downarrow\rangle_i|\!\uparrow\rangle_j
\longrightarrow|\!\downarrow\rangle_i|\!\downarrow\rangle_j$, and
$|\!\downarrow\rangle_i|\!\downarrow\rangle_j
\longrightarrow|\!\downarrow\rangle_i|\!\uparrow\rangle_j$.
A sequence of such conditional two-bit gates supplemented with one-bit rotations constitute a
universal element for QC~\cite{LLOYD}. Usually, a two-bit operation is much slower
than a one-bit operation. Our designs for conditional gates $U_{\rm CPS}$ and $U_{\rm CNOT}$
are {\it efficient} since {\it only one} (instead of two or more)
two-bit operation $U'_{\rm CPS}$ is used.


{\it Persistent currents and entanglement.}~{\bf ---}~The one-bit
circuit modeled by Hamiltonian (1) has two eigenvalues
$E^{(i)}_{\pm}=\pm E_i$, with $E_i=[\varepsilon^2_i(V_{Xi})+
{\overline E}_{Ji}^2]^{1/2}$. The corresponding eigenstates are
$|\psi^{(i)}_+\rangle=\cos\xi_i|\!\uparrow\rangle_i
-\sin\xi_i|\!\downarrow\rangle_i$, and
$|\psi^{(i)}_-\rangle=\sin\xi_i|\!\uparrow\rangle_i
+\cos\xi_i|\!\downarrow\rangle_i$, where $\xi_i=$${1\over
2}\tan^{-1}({\overline E}_{Ji}/\varepsilon_i)$. At these two
eigenstates, the persistent currents through the inductance $L$
are given by $\langle\psi^{(i)}_{\pm}|I|\psi^{(i)}_{\pm}\rangle=
\pm({\overline E}_{Ji}I_{ci}/E_i)\sin(\pi\Phi_e/\Phi_0) +(\pi
LI_{ci}^2/2\Phi_0)\sin(2\pi\Phi_e/\Phi_0)$, where the expansion in
$I$ is retained up to the linear term in $\eta_i$. When a dc SQUID
magnetometer is inductively coupled to the inductance $L$, these
two supercurrents generate different fluxes through the SQUID loop
of the magnetometer and the quantum-state information of the
one-bit structure can be obtained from the measurements. To
perform sensitive measurements with weak dephasing, one could use
the underdamped dc SQUID magnetometer designed previously for the
Josephson phase qubit~\cite{MOOIJ,VAL}.


For the two-bit structure described by Eq.~(2), the Hamiltonian has
four eigenstates and the supercurrents through inductance $L$
take different values at these states. The fluxes produced
by the supercurrents through $L$ can also be detected by the dc SQUID
magnetometer. For instance, when $\varepsilon_k(V_{Xk})=0$ and
${\overline E}_{Jk}>0$ for $k=i$ and $j$, the four eigenstates of the
two-bit structure are
\begin{eqnarray}
|1\rangle &=& {1\over 2}(|\uparrow\rangle_i|\uparrow\rangle_j
-|\uparrow\rangle_i|\downarrow\rangle_j
-|\downarrow\rangle_i|\uparrow\rangle_j
+|\downarrow\rangle_i|\downarrow\rangle_j),\nonumber\\
|2\rangle &=& {1\over 2}(|\uparrow\rangle_i|\uparrow\rangle_j
+|\uparrow\rangle_i|\downarrow\rangle_j
-|\downarrow\rangle_i|\uparrow\rangle_j
-|\downarrow\rangle_i|\downarrow\rangle_j),\nonumber\\
|3\rangle &=& {1\over 2}(|\uparrow\rangle_i|\uparrow\rangle_j
-|\uparrow\rangle_i|\downarrow\rangle_j
+|\downarrow\rangle_i|\uparrow\rangle_j
-|\downarrow\rangle_i|\downarrow\rangle_j),\nonumber\\
|4\rangle &=& {1\over 2}(|\uparrow\rangle_i|\uparrow\rangle_j
+|\uparrow\rangle_i|\downarrow\rangle_j
+|\downarrow\rangle_i|\uparrow\rangle_j
+|\downarrow\rangle_i|\downarrow\rangle_j).\nonumber
\end{eqnarray}
When expansions in $I_i$ and $I_j$ are retained up to the linear terms in
$\eta_i$ and $\eta_j$, the corresponding supercurrents through inductance $L$ are
$\langle k|I|k\rangle={\cal I}_k\sin(\pi\Phi_e/\Phi_0)
+(\pi L{\cal I}_k^2/2\Phi_0)\sin(2\pi\Phi_e/\Phi_0)$ for
$k=1$ to 4, where ${\cal I}_1=-(I_{ci}+I_{cj})$,
${\cal I}_2=I_{cj}-I_{ci}$, ${\cal I}_3=I_{ci}-I_{cj}$,
and ${\cal I}_4=I_{ci}+I_{cj}$. These supercurrents produce different
fluxes threading the SQUID loop of the magnetometer and can be distinguished by
dc SQUID measurements. If the two-bit system is prepared at the
maximally entangled Bell states
$|\Psi^{(\pm)}\rangle=
(|\!\uparrow\rangle_i|\!\downarrow\rangle_j\pm|\!\downarrow\rangle_i
|\!\uparrow\rangle_j)/\sqrt{2}$,
the supercurrents through $L$ are given by
$\langle\Psi^{(\pm)}|I|\Psi^{(\pm)}\rangle=
(\pi L/2\Phi_0)(I_{ci}\pm I_{cj})^2\sin(2\pi\Phi_e/\Phi_0)$.
These two states should be distinguishable by detecting the fluxes (generated by the
supercurrents) through the SQUID loop of the magnetometer.


{\it Discussion.}~{\bf ---}~The
typical switching time $\tau^{(1)}$ during a one-bit operation is of the
order $\hbar/E_J^0$. For the experimental value of $E_J^0\sim 100$~mK,
there is $\tau^{(1)}\sim 0.1$~ns. The switching time $\tau^{(2)}$ for
the two-bit operation is typically of the order
$(\hbar/L)(\Phi_0/\pi E_J^0)^2$. Choosing $E_J^0\sim 100$~mK and
$\tau^{(2)}\sim 10\tau^{(1)}$ (i.e., ten times slower than the one-bit
rotation), we have $L\sim 30$~nH in our proposal, which is experimentally accessible.
A small-size inductance with this value can be made with Josephson junctions.
Our expansion parameter $\eta$ is of the order $\pi^2LE_J^0/\Phi_0^2\sim 0.1$.
Our inductance $L$ is related with the inductance $L'$ in ~\cite{MSS, ASZ} by
$L'=(C_J/C_{qb})^2L$. Let us now consider the case when $\tau^{(2)}\sim 10\tau^{(1)}$.
For the earlier design~\cite{ASZ}, $C_J\sim 11C_{qb}$ since $C_g/C_J\sim 0.1$,
which requires the inductance $\sim 3.6$~$\mu$H. Such a large inductance is
difficult to fabricate at nanometer scales. In the improved design~\cite{MSS},
$C_J\sim 2C_{qb}$, greatly reducing the inductance to $\sim 120$ nH.
This inductance is about four times larger than the one used in our scheme.


All charge qubits suffer decoherence due to the fluctuations of voltage sources
and fluxes. Ref.~\cite{MSS} shows that the gate voltage
fluctuations play the dominant role in producing decoherence.
The estimated dephasing time is $\tau_{\varphi}\sim 10^{-4}$~s, allowing
in principle $10^6$ coherent single-bit manipulations. When a probe junction
is used for measurements, the experimental observations of coherent oscillations
in the Josephson charge qubits show that the phase coherence time is only
about $2$ ns~\cite{NPT,NPYT}. In this experimental setup, background charge
fluctuations and the probe-junction measurement may be two of the major factors in
producing decoherences. Though the charge fluctuations are important only in
the low-frequency region and can be reduced by the echo technique~\cite{NPYT} and
by shifting the gate voltage to the degeneracy point, an effective technique for
suppressing charge fluctuations still needs to be explored. As for the measurement,
it has also been a challenge to design effective detecting devices.


In conclusion, we propose a {\it scalable} quantum computer with Josephson charge
qubits. We employ a common inductance to couple all charge qubits and design
{\it switchable interbit couplings} using two dc SQUIDs to connect the island in each
Cooper-pair box. The proposed QC architectures are scalable since any two charge
qubits can be effectively coupled by an experimentally accessible inductance.
Furthermore, we formulate an {efficient} QC scheme in which {\it only one} two-bit
operation is used in the conditional transformations such as
controlled-phase-shift and controlled-NOT gates.

We thank Yu.~Pashkin for useful discussions.
We acknowledge support from the US ARDA, AFOSR,
and the US National Science Foundation grant No.~EIA-0130383.


\begin{references}
\bibitem{eqn1} The Hamiltonian of the one-bit circuit shown in Fig.~2(a) is
$H=H_i+{1\over 2}LI_i^2$, with
$H_i = E_{ci}(n_i-n_{Xi})^2-2E_{Ji}(\Phi_{Xi})\cos(\pi\Phi/\Phi_0)\cos\varphi_i$.
%
Here, the phase $\varphi_i=(\phi_{iA}+\phi_{iB})/2$ is canonically conjugate
with the number of the extra Cooper pairs on the island and the persistent
circulating current $I_i$ in the superconducting loop is given by
$I_i=2I_{ci}\cos\varphi_i\sin(\pi\Phi_e/\Phi_0+\pi LI_i/\Phi_0)$,
where $I_{ci}=-\pi E_{Ji}(\Phi_{Xi})/\Phi_0$.
%
In the spin-${1\over 2}$ representation, based on charge states
$|\!\uparrow\rangle_i=|n_i\rangle$ and $|\!\downarrow\rangle_i=|n_i+1\rangle$,
the reduced Hamiltonian of the system becomes Eq.~(1),
%
where $\varepsilon_i(V_{Xi})={1\over
2}E_{ci}[C_iV_{Xi}/e-(2n_i+1)]$, and
$\overline{E}_{Ji}(\Phi_{Xi},\Phi_e,L)\,=\,
E_{Ji}(\Phi_{Xi})(\gamma_i+\eta_i\alpha_i\beta_i)$.
%
Here $\alpha_i$, $\beta_i$, and $\gamma_i$ are power series of the
expansion parameter $\eta_i=\pi LI_{ci}/\Phi_0$, which is $\sim
0.1$ in our case (see the discussion part above). These have the
same expressions as $\alpha$, $\beta$, and $\gamma$ in
Ref.~\cite{YLZ}, with $I_c$ and $\Phi_X$ there replaced by
$I_{ci}$ and $\Phi_e$. Retained up to second-order terms in the
expansion parameter,
$\overline{E}_{Ji}(\Phi_{Xi},\Phi_e,L)\,=
\,E_{Ji}(\Phi_{Xi})\cos(\pi\Phi_e/\Phi_0)\xi$,
with $\xi=1-{1\over 2}\eta_i^2\sin^2(\pi\Phi_e/\Phi_0)$.
\bibitem{eqn2} The Hamiltonian of the two-bit circuit given in Fig.~2(b) is
$H=H_i+H_j+{1\over 2}L(I_i+I_j)^2$, where
$I_i=2I_{ci}\cos\varphi_i\sin[\pi\Phi_e/\Phi_0+\pi L(I_i+I_j)/\Phi_0]$
is the persistent circulating current contributed by the Cooper-pair box $i$.
%
Interchanging $i$ and $j$ in $I_i$ gives the circulating
current $I_j$ contributed by the Cooper-pair box $j$. In the spin-${1\over 2}$
representation, the Hamiltonian of the system reduces to Eq.~(2),
%
where the intrabit coupling $\overline{E}_{Jk}$
and the interbit coupling $\Pi_{ij}$ are also power series,
which have the same expressions as in Ref.~\cite{YLZ},
but with:
$L_i$, $L_j$, and $M_{ij}$ replaced by $L$;
$\Phi_{Xi}$ and $\Phi_{Xj}$ replaced by $\Phi_e$;
and $E^0_{Jk}$ by $E_{Jk}(\Phi_{Xk})$.
%
Up to second-order terms,
$\overline{E}_{Ji}(\Phi_{Xi},\Phi_e,L)\,=\,E_{Ji}(\Phi_{Xi})\cos(\pi\Phi_e/\Phi_0)\xi$,
with $\xi=1-{1\over
2}(\eta_i^2+3\eta_j^2)\sin^2(\pi\Phi_e/\Phi_0)$, and
$\Pi_{ij}=-LI_{ci}I_{cj}\sin^2(\pi\Phi_e/\Phi_0)$.
%
When the two qubits are far apart, the inductance of the wires connecting them
might not be neglected.  However, the reduced two-bit Hamiltonian is still
given by Eq.~(2), but with $\overline{E}_{Jk}$ and $\Pi_{ij}$ slightly modified.
%
In deriving the interbit coupling $\Pi_{ij}$, here
we ignore the collective $LC$-oscillator mode associated with the total charge $q$
accumulated on the gate capacitors of the arrays of Cooper-pair boxes~\cite{MSS}.
%
When it is taken into account, a term $(q-\sum_k 2en_k\chi_k)^2/2C_{\Sigma}$,
with $C_{\Sigma}=\sum_k 2C_J(C_k+2C_J)/(C_k+4C_J)$ and $\chi_{k}=C_k/(C_k+4C_J)$,
needs to be included in the Hamiltonian. Fast collective oscillations in the $LC$ circuit
then yield an effective coupling between qubits $i$ and $j$, and the effective coupling
strength $|\Lambda_{ij}|$ is $\sim \chi_i\chi_j|\Pi_{ij}|$.
For typical values of $C_k/C_J\sim 0.1$, there is $\chi_k\sim 0.025$.
Thus, we have $|\Lambda_{ij}|\sim 6\times 10^{-4}|\Pi_{ij}|$, implying that neglecting
the $LC$-oscillator effect here is quite reasonable.
\end{references}
\end{document}